\documentclass[conference]{IEEEtran}
\usepackage{bm,cite,algorithm,algorithmic,float,amsmath,amssymb}

\usepackage{amssymb}
\usepackage{amsmath}
\usepackage{graphicx}
\usepackage{cite}
\usepackage{citesort}
\usepackage{balance}
\usepackage[utf8x]{inputenc}
\usepackage{amsthm}
\usepackage{framed}

\IEEEoverridecommandlockouts

\usepackage{graphicx,epstopdf}
\usepackage{epsfig}	
\usepackage{amsfonts,balance}
\usepackage{bbm}
\floatname{algorithm}{Algorithm}

\begin{document}

\title{The Unreasonable Effectiveness of Blood Pressure Measurement: Molecular Communication in Biological Systems}

\author{\authorblockN{Malcolm Egan$^{\dag}$, Adam Noel$^{\ddag\ddag}$, Yansha Deng$^\ddag$, Maged Elkashlan$^{\ddag}$ and Trung Q. Duong$^{\ast}$
\authorblockA{\authorrefmark{2}\footnotesize Faculty of Electrical Engineering, Czech Technical University in Prague, Czech Republic }
\authorblockA{\authorrefmark{9}\footnotesize Department of Electrical and Computer Engineering, University of British Columbia, CA }
\authorblockA{\authorrefmark{3}\footnotesize School of Electronic Engineering and Computer Science, Queen Mary University of London, UK }
\authorblockA{\authorrefmark{1}\footnotesize School of Electronic Engineering and Computer Science, Queen's University Belfast, UK }
}}

\maketitle

\begin{abstract}
Arterial blood pressure is a key vital sign for the health of the human body. As such, accurate and reproducible measurement techniques are necessary for successful diagnosis. Blood pressure measurement is an example of molecular communication in regulated biological systems. In general, communication in regulated biological systems is difficult because the act of encoding information about the state of the system can corrupt the message itself. In this paper, we propose three strategies to cope with this problem to facilitate reliable molecular communication links: \textit{communicate from the outskirts}; \textit{build it in}; and \textit{leave a small footprint}. Our strategies---inspired by communication in natural biological systems---provide a classification to guide the design of molecular communication mechanisms in synthetic biological systems. We illustrate our classification using examples of the first two strategies in natural systems. We then consider a molecular link within a model based on the Michaelis-Menten kinetics. In particular, we compute the capacity of the link, which reveals the potential of communicating using our \textit{leave a small footprint} strategy. This provides a way of identifying whether the molecular link can be improved without affecting the function, and a guide to the design of synthetic biological systems.
\end{abstract}

\maketitle

\section{Introduction}

Hypertension---commonly known as high arterial blood pressure---is a key risk factor for stroke, coronary heart disease, congestive heart failure, renal failure, and peripheral vascular disease \cite{Perloff1993}. As such, accurate and reproducible techniques to measure blood pressure are necessary for successful medical diagnoses. To this end, a non-invasive technique known as sphygmomanometry---simply using an inflatable cuff around a patient's arm---has been developed, which is easy to perform and sufficiently accurate for the purposes of diagnosis.

It is surprising that sphygmomanometry is accurate. Arterial blood pressure is highly regulated; when blood pressure becomes too high or too low there are severe consequences. In sphygmomanometry, arterial blood flow in the arm is temporarily cut off (or occluded) \cite{Perloff1993}, which means there is a local change in blood pressure. Naively, this local change would be expected to cause the regulatory system to rapidly adjust the blood flow in the rest of the body, which would invalidate the measurement. Fortunately, the special structure of the arterial blood pressure regulatory system prevents this, facilitating accurate and reproducible measurement.

Blood pressure measurement can be viewed as a molecular communication problem, where the network of arteries is the source and the receiver is the sphygmomanometer. This is because arterial blood pressure is related to the flow of the molecules that comprise blood.

In fact, communication in biological systems is closely related to measurement; typically, the goal is to communicate the state---often the concentration of a compound---of one component to another. In contrast with many popular molecular communication links based on models of isolated chemical systems \cite{Srinivas2012,Egan2014,Nakano2013,Nakano2012}, communication in biological systems is complicated by the presence of feedback mechanisms, which regulate the concentration of the molecules within the system.

As encoding the concentration of a given compound for communication over a molecular link usually requires a reaction involving the compound, it means that the concentration of the compound will be altered. This can then initiate a feedback mechanism, which will change the concentration of the compound within the component. That is, simply by encoding the concentration, the message itself will be corrupted---the key difficulty of communication in biological systems.

In this paper, we propose three general strategies inspired by nature to design reliable synthetic molecular communication links in biological systems; both in synthetic biological systems, and to improve biological function in existing systems. Importantly, our strategies form a classification of different communication mechanisms in biological systems.

Our first strategy, \textit{communicate from the outskirts}, is inspired by blood pressure measurement in the human body. As we detail in Section~\ref{sec:first_strategy}, sphygmomanometry is possible because the baroreceptors---internal blood pressure sensors---are localized near the neck and heart \cite{Williamson1994}. As such, local changes in blood flow far away from the neck and heart will not be immediately detected. More generally, this suggests that biological components that are regulated by loosely coupled reactions can support reliable molecular communication.

Our second strategy, \textit{build it in}, applies to biological systems with regulation formed by tightly coupled reactions, which means that the \textit{communicate from the outskirts} strategy cannot be applied. In the \textit{build it in} strategy, which is detailed in Section~\ref{sec:second_strategy}, the component transmits information at the same time that molecules of the regulated compound are produced via a communication mechanism that does not interfere with the regulation of the system.

An important example in nature of the \textit{build it in} strategy is cell-to-cell communication within bacteria colonies, often known as quorum sensing \cite{Melke2010}. The key drawback of this strategy is that the biological system is required to be more sophisticated---relative to systems that do not couple feedback and communication. As such, design and operation costs (e.g., energy consumption) are much higher. Despite this drawback, new techniques from synthetic biology mean that the strategy is in principle able to be adopted to implement synthetic molecular links in biological systems \cite{Khalil2010}.

Our final strategy, \textit{leave a small footprint}, is targeted at biological systems that do not support the first two strategies; i.e., the regulation is strongly coupled, or the costs do not justify building in the communication mechanism. The basic principle of this strategy (in Section~\ref{sec:third_strategy}) is that the concentration of a compound can be communicated, as long as the perturbation of the concentration---due to the encoding process---is bounded. Although this means that the capacity of the link is constrained, the function of the biological system can be preserved.

Next, we evaluate in Section~\ref{sec:foot} the potential and limitations of our \textit{leave a small footprint} strategy in a biochemical system with Michaelis-Menten kinetics---widely used to model natural biological systems regulated by enzymes \cite{Reed2009,Noel2014}. In particular, we compute the capacity of the link, which provides a way of identifying whether the molecular link in natural biological systems can be improved without affecting the function, and a guide to the design of synthetic biological systems.

\section{Three Communication Strategies}

The concentrations of key chemical compounds in biological systems often depend on several coupled chemical reactions, which are regulated by complex feedback mechanisms. As such, even small changes in the concentration of a single compound can affect the concentration of other compounds within the system. This means that reliable molecular communication is not possible without carefully accounting for how the encoding and transmission processes interact with the feedback mechanisms. In sharp contrast with isolated chemical systems (such as the standard molecular timing channel \cite{Srinivas2012,Egan2014}), the interaction of the communication mechanism with the whole system must be considered, not simply the direct communication channel.

In this section, we propose three strategies inspired by nature for reliable molecular communication in biological systems: \textit{communicate from the outskirts}; \textit{build it in}; and \textit{leave a small footprint}. Here, reliability is not just the probability of error: the effect of the communication mechanism on the function of the biological system must also be captured. To ensure that the biological system can preserve its function, we identify scenarios where each strategy is best suited and can be adopted to design synthetic molecular communication links.

There are two key factors that determine the applicability of each strategy: reaction coupling; and cost constraints. The reaction coupling is a measure of the interactions between different chemical compounds. On the other hand, the cost constraints correspond to energy consumption or complexity of the system. We summarize our strategies with a coarse classification of the required reaction couplings and cost constraints in Table~\ref{table:strategy_comp}. Observe that \textit{communicate from the outskirts} is applicable irrespective of the cost constraint, as long as there is weak coupling between reactions. On the other hand, the other two strategies work irrespective of the reaction coupling; instead, they depend on whether there is a high (abundant resources) or low (limited resources) cost constraint.

\begin{table}[!h]
\begin{center}
\caption{Summary of strategies and their applicability.}\label{table:strategy_comp}
 \begin{tabular}{|c|c|c|}
 \hline
 Strategy & Reaction Coupling & Cost Constraints \\ \hline
 \textit{Communicate from the outskirts}     & Weak & Low/High \\
 \hline
 \textit{Build it in} & Weak/Strong & High \\
 \hline
 \textit{Leave a small footprint} & Weak/Strong & Low \\
 \hline
 \end{tabular}
\end{center}

\end{table}

\subsection{Communicate From the Outskirts}\label{sec:first_strategy}

Our first strategy, \textit{communicate from the outskirts}, exploits situations where chemical compounds carrying useful information do not strongly interact with other parts of the system. As such, the concentration of these compounds can be perturbed by the communication link without interacting with any feedback mechanisms, which would corrupt the message to be transmitted.

A key scenario where this strategy occurs in nature (i.e., without synthetically modifying the biological system) is in arterial blood pressure measurement. In this case, the goal is to communicate the arterial blood pressure to outside the body. Using sphygmomanometry---with an inflatable cuff---this is achieved by applying external pressure and cutting off blood flow in the artery through the arm of the patient, and slowly releasing the external pressure. Blood pressure can then be measured by observing the Karotkoff sounds (see \cite{Perloff1993} for a full description of the procedure).

As it is important for bodily function to keep blood pressure nearly constant (i.e., in homeostasis), cutting off blood flow would be expected to alter the blood pressure in the rest of body---making an accurate measurement of the blood pressure impossible. However, this is not the case. As the baroreceptors---which regulate blood pressure---are predominately located in arteries near the neck and the heart, it takes up to half an hour \cite{Williamson1994} for the local change in blood pressure in a limb to cause a significant change in blood pressure where the baroreceptors are located.

The slow change in blood pressure near the baroreceptors means that it is possible to perturb the blood flow in the arm without affecting the blood flow in the rest of body. This facilitates accurate measurement, and hence reliable communication of blood pressure from within the body to the outside. The key feature that enables reliable measurement is the loose coupling between the local blood pressure in the arm and the blood pressure in the rest of the body.

Another scenario where the \textit{communicate from the outskirts} strategy can be applied is in biological systems regulated by enzymes that only react with a small number of chemical compounds. This weak coupling means that the concentration of enzymes can be altered without affecting the rest of the biological system. As such, reliable communication via other carrier molecules of the enzyme concentrations (and the compounds they react with) can be possible, while still preserving the overall function of the system.

Importantly, natural biological systems where enzyme reactions are weakly coupled occur frequently. The reason for this is that these systems are approximately scale-free, which means that enzymes catalyzing several reactions occur rarely. This suggests that the function of the system can be maintained, even after perturbing a small number of the reactions, as long as the reactions are not crucial to the function of the system. As scale-free networks in biological systems are conjectured to be widespread \cite{Albert2005}, there are a range of possible applications of the \textit{communicate from the outskirts} strategy; including in the bacteria \textit{E. Coli} and the eukaryote \textit{C. elegans} \cite{Jeong2000}.

\subsection{Build It In}\label{sec:second_strategy}

Our second strategy, \textit{build it in}, facilitates reliable communication by coupling the communication mechanism to the regulatory system. That is, a communication link is induced by the concentration of one or more chemical compounds, \textit{without} altering any of the concentrations. This is highly desirable; however, it also increases the complexity, and as such the cost of constructing and operating the biological system.

Despite the increased complexity, the \textit{build it in} strategy naturally occurs in cell-to-cell communication in bacteria colonies \cite{Melke2010,Einolghozati2012}; known as quorum sensing. The most popular examples of quorum sensing are in the bacteria \textit{Vibrio fischeri} or \textit{Vibrio harveyi} \cite{Miller2001}, where the communication link is induced by the production of a compound known as an autoinducer. Each bacterium can then detect the total concentration of the autoinducer, which initiates a positive feedback mechanism---leading to the emission of light (see \cite{Einolghozati2012} for more details). The intensity of the light can then reveals useful properties of the bacteria colony, such as the number of bacterium.

In principle, the \textit{build it in} strategy can be applied to any system regulated via DNA transcription \cite{Reed2009} by modifying via synthetic biology techniques the transcription process so that a signalling molecule is emitted, with the amount dependent on the concentration of the regulated compound. The key advantage is that the communication mechanism does not interfere with the message; that is, the concentration of the compound to observed. In the case of quorum sensing, this is achieved by signalling with light, which does not affect the concentration of autoinducer. The drawback is that this approach is costly in terms of both construction and operation, and requires careful design to ensure that the biological system retains its intended function. This is particularly important in synthetic biological systems with strict energy consumption constraints; due to limited resources for biochemical energy production mechanisms.

\subsection{Leave a Small Footprint}\label{sec:third_strategy}

Our final strategy, \textit{leave a small footprint}, is an approach of last resort, used when neither of the other strategies are suitable due to the structure or the cost of modifying the biological system. Our previous two strategies exploited the structure of biological systems to ensure that the communication mechanism does not perturb the concentration of key chemical compounds. Unfortunately, when perturbation is unavoidable, the only course of action is to limit the perturbation. This means that communication is possible---albeit at a reduced rate---while ensuring that the function of the biological system is preserved.

In order to apply the \textit{leave a small footprint} strategy, both the communication link (which determines the rate) and the biochemical reaction network (which determines the perturbation) need to be considered. It is worth noting that this is in contrast to isolated chemical systems---such as the molecular timing channel \cite{Srinivas2012}---which only consider the communication link.

To explore the reliability of the \textit{leave a small footprint} strategy, we develop a concrete example of a biochemical reaction network enclosed within a permeable membrane. The goal is to communicate the concentration of a key compound within the membrane to the outside. This model and its properties are developed in the next section.

\section{Reliable Communication With A Small Footprint}\label{sec:foot}

\subsection{System Model}\label{sec:sys_model}

Consider a biological system enclosed by a membrane, with volume $V$. Inside the membrane is a biochemical reaction network, which involves a chemical species $S$. We desire to communicate the concentration of $S$ outside the membrane. As the membrane is impermeable to $S$, this can be achieved by introducing an enzyme $E$, which initiates the reaction
\begin{align}\label{eq:reaction_SP}
E + S \overset{k_1}{\underset{k_{-1}}{\rightleftarrows}} \overline{ES} \overset{k_2}{\rightarrow} E + P,
\end{align}
where the parameters $k_1,k_{-1}$, and $k_2$ are reaction rates.

\begin{figure}[h]
\centering{\includegraphics[width=60mm]{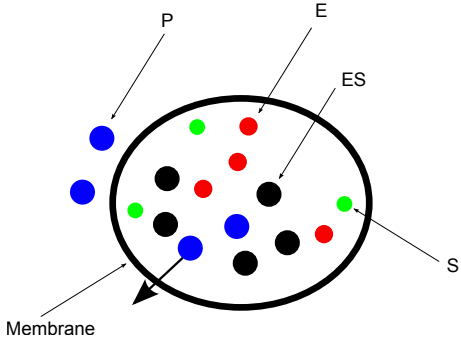}}
\caption{Illustration of the biological system component with Michaelis-Menten kinetics.}\label{fig:m_m_sys}
\end{figure}

We assume that the membrane is impermeable to $S$ and the complex $\overline{ES}$. Importantly, we also assume the enzyme $E$ can be introduced or removed at any time, which controls the reduction of $S$ within the membrane. As such, the reduction in the concentration of $S$ can be constrained, which preserves the function of the original biological system within the membrane (i.e., without $E$). The system directly involved in producing $P$ then consists of $E$, $S$, $\overline{ES}$ and $P$, which is illustrated in Fig.~\ref{fig:m_m_sys}.

As the membrane is impermeable to $S$ and $\overline{ES}$, the concentration of $S$ is communicated outside the membrane by $P$. In order to support reliable communication, we aim to produce the maximum concentration of $P$ while ensuring that the concentration of $S$ is not significantly reduced.

The reaction in (\ref{eq:reaction_SP}) determining the concentrations of $S$ and $P$ is modeled by Michaelis-Menten kinetics \cite{Reed2009}, which is known to govern a range of biochemical reactions involving enzymes. In particular, the concentrations of each compound at time $t$, $[E](t),[S](t),[\overline{ES}](t),[P](t)$ respectively, are governed by coupled differential equations, which can be derived assuming mass-action kinetics \cite{Chellaboina2009}. In particular, the differential equations are given by
\begin{align}\label{eq:ODEs}
\frac{\partial [S](t)}{\partial t} &= -k_1[E](t)[S](t) + k_{-1}[\overline{ES}](t),\notag\\
\frac{\partial [E](t)}{\partial t} &= -k_1[E](t)[S](t) + (k_{-1} + k_2)[\overline{ES}](t),\notag\\
\frac{\partial [\overline{ES}](t)}{\partial t} &= k_1[E](t)[S](t) - (k_{-1} + k_2)[\overline{ES}](t).
\end{align}

The initial concentrations at $t = 0$ are $[S](0) = S_0$, $[E](0) = E_0$, $[\overline{ES}](0) = \overline{ES}_0$, and $[P](0) = 0$. We also assume that the following conservation law holds
\begin{align}
[E](t) + [\overline{ES}](t) = [E](0) + [\overline{ES}](0) = E_T.
\end{align}

Often, the initial concentration of $S$, given by $S_0$, has a finite number, $K$, of possible values; i.e., $S_0 \in \{s_1,\ldots,s_K\}$. This is due to the fact that in many biological systems $S_0$ will be determined by other reactions, which we do not explicitly model. In particular, the finite number of states can arise due to the presence of a positive feedback network; a key example is the biochemical switch exploited by bacteria colonies \cite{Melke2010}.

We now turn to the model of the molecular communication link, which operates using discrete time slots of duration $T$. At the beginning of each time slot, the biochemical switch is activated by introducing $E$, which begins production of $P$. When a molecule of $P$ is produced, we assume that the membrane it passes through is an erasure channel with success probability $q$. The success probability captures the loss of molecules during diffusion and the ability for the receiver to detect $P$. We note that $q$ can in principle be computed from the underlying stochastic diffusion process. This means that the received signal on the outside of the component membrane $Y_j$ in time slot $j$ is given by
\begin{align}\label{eq:bin_channel}
Y_j = \sum_{i=1}^{N_j} X_{i,j},
\end{align}
where $N_j$ is the number of molecules of $P$, which lies between $[0,N_{\max}]$ with $N_{\max}$ the maximum number of molecules of $P$ that can be produced in a time slot. The variable $X_{i,j}$ is then a Bernoulli random variable for the $i$-th molecule with success probability $q$.

To ensure that the biological system can still perform its function, we introduce a perturbation constraint on $[S](t)$, which determines $N_{\max}$. In particular, the following condition must hold
\begin{align}\label{eq:pert_cond}
S_0 - [S](T) \leq \Delta,
\end{align}
where $\Delta$ is the maximum perturbation. The perturbation constraint can be enforced by removing the enzyme $E$ from the system after time $T$, which halts the reaction in (\ref{eq:reaction_SP}) and flushes $P$ from the system. This means that the concentration of $S$ returns to its initial state, $S_0$, which occurs due to the feedback mechanism regulating $[S]$.

The key consequence of the perturbation constraint is that the maximum concentration, $[P](t^*)$ (equivalent to $N_{\max}/V$), of $P$ is bounded. As we will show next, this has important consequences for the capacity of the molecular communication link through the membrane.

\subsection{Capacity Analysis}


We now turn to the capacity of the molecular link in our model. Recapping, the input is the concentration of $P$, and the membrane forms an erasure channel for each molecule of $P$. Moreover, the input concentration $[P]$ must be bounded so that the condition in (\ref{eq:pert_cond}) is satisfied.

It is worth pointing out that the link capacity does in fact have physical meaning as a measure of reliability in our model. This is because it is in principle possible to perform coding on the input message $[P]$ via another chemical reaction network, which would need to be introduced on top of our model. As a general theory of coding using reaction networks in molecular communications has not yet been developed, we do not pursue this further here.

In order to compute the capacity, we first need to obtain an upper bound on $[P](t)$ so that the perturbation constraint (\ref{eq:pert_cond}) holds. This is achieved by computing the time
\begin{align}
t^* = \sup \{t: S_0 - [S](t) \leq \Delta\}.
\end{align}

In general, obtaining a simple solution to the differential equations in (\ref{eq:ODEs}) is not straightforward. Fortunately, we can assume the pseudo-steady state (PSS) condition holds \cite{Schnell1997}, which is guaranteed when $t_S \ll t_{\overline{ES}}$ (which are characteristic time scales, given by the inverses of the reaction rates), or $E_0 \ll K_m + S_0$, where \cite{Schnell1997}
\begin{align}
K_m = \frac{k_{-1} + k_2}{k_1}.
\end{align}
Importantly, these conditions hold in a wide range of scenarios.

Under the PSS condition, the concentration for the substrate $S$ can be written as
\begin{align}
\frac{\partial [S](t)}{\partial t} &= -\frac{k_2E_T[S](t)}{K_m + [S](t)}.
\end{align}
The concentration of $S$ can then be written as \cite{Schnell1997}
\begin{align}\label{eq:S_dyn}
[S](t) = K_mW\left(F(t)\right),
\end{align}
where $W(\cdot)$ is the Lambert W-function, and
\begin{align}
F(t) = \frac{S_0}{K_M} \exp\left(\frac{1}{K_M}\left(S_0 - V_{\max}t\right)\right).
\end{align}

Using (\ref{eq:S_dyn}), the time $t^*$ can be obtained numerically by solving $[S](t) = S_0 - \Delta$, for $t$. We also note that the concentration of $P$ can be obtained by solving \cite{Murray2002}
\begin{align}\label{eq:P_dyn}
\frac{\partial [P](t)}{\partial t} = \frac{V_{\max}[S](t)}{K_m + [S](t)},
\end{align}
where $V_{\max} = k_2E_T$.

Next, we derive the maximum value of $[P](t)$ such that the perturbation constraint holds, which is denoted by $[P](t^*)$. This means that the input concentration must lie in $[0,[P](t^*)]$. From (\ref{eq:P_dyn}), the maximum input concentration is given by
\begin{align}\label{eq:P_max}
[P](t^*) &= \int_0^{t^*} \frac{V_{\max}[S](\tau)}{K_m + [S](\tau)}d\tau.
\end{align}

In general, $[P](t^*)$ must be solved numerically. Fortunately, when $S_0 - \Delta \gg K_m$, we can obtain an accurate approximation for $[P](t^*)$ in closed-form. Note that this occurs when $k_1 \gg k_2 + k_{-1}$, which holds, for instance, in mitochondrial fumarase. This is an important enzyme involved in the Krebs cycle---a key biochemical energy production mechanism \cite{Matthews1999}.

Next, observe that
\begin{align}\label{eq:p_conc_bound}
[P](t^*) &= \int_0^{t^*} \frac{V_{\max}[S](\tau)}{K_m + [S](\tau)}d\tau\notag\\
&\leq \int_0^{t^*} V_{\max}d\tau = t^*V_{\max},
\end{align}
which is tight when $K_m$ is small compared with $[S](t),~\forall t < t^*$. Now, the Lambert W-function can be approximated as \cite{Borsch1961}
\begin{align}
W(e^{x + a}) \approx x\left(1 - \frac{\log x - a}{x + 1}\right),
\end{align}
when $x \gg a$. Applying this approximation to (\ref{eq:S_dyn}), we obtain
\begin{align}
[S](t) &\approx (S_0 - V_{\max}t)\left(1 - \frac{\log\left(\frac{S_0 - V_{\max}t}{S_0}\right)}{\frac{S_0 - V_{\max}t}{K_m} + 1}\right)\notag\\
&\approx S_0 - V_{\max}t,
\end{align}
when $K_m$ is small (such as in mitochondrial fumarase). This means that
\begin{align}
V_{\max}t^* \approx S_0 - [S](t^*) = \Delta.
\end{align}
After substituting into (\ref{eq:p_conc_bound}), we obtain
\begin{align}\label{eq:Ptstar_approx}
[P](t^*) \approx \Delta.
\end{align}
That is, the maximum concentration of $P$ such that the perturbation constraint holds is approximately equal to the maximum perturbation. Moreover, $T = t^*$ is an appropriate choice for the duration of a time slot. We point out that this result is intuitive as any loss of $S$ should eventually be converted to $P$, which is interesting as this conversion occurs very rapidly (within time $t^*$).

Now that we have the maximum input concentration of $P$---corresponding to $N_j$ molecules---we turn to the erasure channel, which models the communication channel through the membrane. We note that the received signal $Y_j$ is binomially distributed with parameter $q$, conditioned on $N_j = V[P](T)$ ($V$ is the volume of the system and $T = t^*$).

It is worth noting the similarities between the communication link in our model and that of the quorum sensing model in \cite{Einolghozati2012}. The cause of the similarity is that the underlying channel in both models is the binomial channel in (\ref{eq:bin_channel}), induced by an erasure channel acting on each molecule. Despite the similarities, there are two key differences. The first is the maximum input concentration. In the quorum sensing model, this is determined by the autoinducer production mechanism in each bacterium, and the number of bacteria; while in our model, the maximum input concentration is determined by the perturbation constraint.

Second, the message in our model is $N_j$, the number of molecules of $P$; while in the quorum sensing model, the message is the concentration of autoinducer, which behaves similarly to the erasure probability $q$ in (\ref{eq:gauss_channel}) rather than $N_j$. To see this, note that the central limit theorem implies that the received signal, conditioned on $N_j$, can be approximated by
\begin{align}\label{eq:gauss_channel}
\tilde{Y_j} \sim \mathcal{N}(\mu,\sigma^2),
\end{align}
for sufficiently large $N_j$, where $\mu = N_jq$ and $\sigma^2 = N_j^2q^2(1-q)^2$.

Finally, we compute the capacity using the Blahut-Arimoto algorithm \cite{Blahut1972}, where the input is bounded in $[0,[P](T)]$ (again, $T = t^*$) and the channel is given by (\ref{eq:bin_channel}). To demonstrate the effect of the perturbation constraint, we provide detailed numerical results in Section~\ref{sec:num_results}.

\subsection{Numerical Results}\label{sec:num_results}

In this section, we investigate the effect of the perturbation constraint and the channel erasure probability on the capacity.

First, we demonstrate that (\ref{eq:Ptstar_approx}) is an accurate approximation for $[P](t^*)$ for small $K_m$. Observe from Table~\ref{table:setup} that the perturbation constraint $\Delta$ and the maximum concentration of $P$, $[P](t^*)$ are in precise agreement. This result is convenient as it provides a simple closed-form expression for the integral in (\ref{eq:P_max}), which would otherwise require a root-finding method (to obtain $t^*$) and numerical integration.

\begin{table}[!h]
\caption{A comparison of the perturbation constraint, $\Delta$, and the maximum concentration of $P$, $[P](T)$ (from (\ref{eq:P_max}), with $T = t^*$). Parameters are: $K_m = 0.1$; $S_0 = 30$; and $V_{\max} = 1$.}\label{table:setup}
\begin{center}
 \begin{tabular}{|c|c|c|c|c|c|}
 \hline
 $\Delta$ & 0.1 & 0.2 & 0.3 & 0.4 & 0.5 \\ \hline
 $[P](t^*)$     & 0.10 & 0.20 & 0.30 & 0.40 & 0.50 \\
 \hline
 \end{tabular}
\end{center}
\end{table}

\begin{figure}[h!]
\centering{\includegraphics[width=90mm]{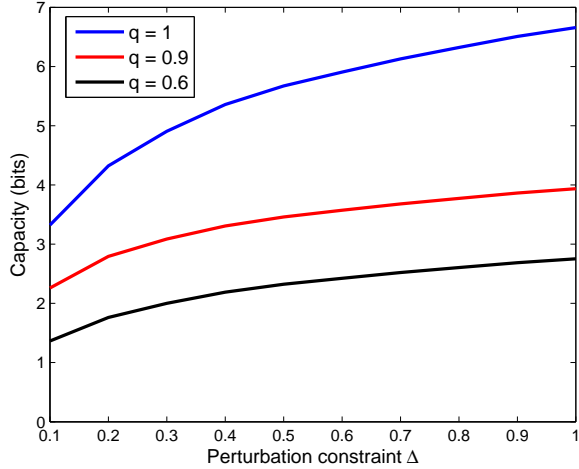}}
\caption{Plot of capacity as the perturbation constraint, $\Delta$, varies. Parameters are: $K_m = 0.1$; $S_0 = 10$; $V_{\max} = 1$; and the system volume is $V = 100$.}\label{fig:q_effect}
\end{figure}

Next, we turn to the effect of $\Delta$ and the erasure probability $q$ on the capacity. Fig.~\ref{fig:q_effect} plots the capacity (in bits) for varying $\Delta$ and $q$. As expected, the capacity increases when $\Delta$ increases as it leads to a higher value of $[P](t^*)$, and also when $q$ increases as the membrane channel has lower noise. For low $\Delta$, the capacity increases rapidly; while for higher $\Delta$ (i.e., $\Delta > 0.1$), the slope decreases. This suggests an analogy with the Gaussian channel, where $\Delta$ plays the same role as the transmit power (due to the intimate relationship with $[P](t^*)$). In particular, there are diminishing gains as $\Delta$ increases. Importantly, the parameters $S_0$ and $V_{\max}$ do not play an important role with this choice of $K_m$ due to (\ref{eq:Ptstar_approx}).

\section{Conclusions}

A popular approach for modeling of molecular communication systems is to assume an isolated chemical system. Unfortunately, this approach cannot be applied for biological system-based applications, as the concentrations of compounds are often coupled to complex feedback mechanisms. As such, it is important to develop strategies to ensure that reliable communication is possible, without initiating the feedback loops. This is necessary as biological systems are often required to be highly stable, and cannot fulfil their function when the system is perturbed by large changes in compound concentrations.

To this end, we have introduced three general strategies for molecular communications in biological systems: \textit{communicate from the outskirts}; \textit{build it in}; and \textit{leave a small footprint}. Our strategies form a classification of communication mechanisms applicable in biological systems, with each strategy appropriate in different types of systems. We have shown that existing work has fallen under the first two strategies, and detailed new scenarios where these strategies can be applied.

To investigate the reliability of systems exploiting the \textit{leave a small footprint} strategy, we introduced a model based on Michaelis-Menten kinetics, which is widely used to model enzyme biochemical reactions. With our model, we demonstrated how the capacity is affected by the perturbation constraint and the molecular channel.

Designing reliable molecular communication mechanisms in biological systems remains a difficult problem. We expect that using our three strategies will help identify the right techniques to ensure that biological systems retain their function, while facilitating reliable communication.


\bibliographystyle{ieeetr}
\bibliography{icc_molecular_desktop}

\end{document}